\documentclass[12pt]{article}
\usepackage{epsfig}
\usepackage{amsfonts}
\usepackage{latexsym}
\usepackage{amsmath,amssymb}
\usepackage{verbatim}
\usepackage{setspace}

\usepackage[textheight=9in, textwidth=6.5in, letterpaper]{geometry}
\def\half{{1\over 2}}
\numberwithin{equation}{section}

\def\ip{${\mathcal I}^+$}

\def\e{{\epsilon}}
\def\cs{{\cal S}}

\def\Psz{\Psi^{0} }
 \def\p{\partial}
 \def\bz{{\bar z}}
 
\def\0{{(0)}}
\def\1{{(1)}}
\def\2{{(2)}}
 \def\cL{{\cal L}}

\def\ci{{\mathcal I}}
\def\ipp{${\mathcal I}^+_+$}
\def\<{\langle }
\def\>{\rangle }
\def\bw{{\bar w}}
\newcommand{\bea}{\begin{eqnarray}}
\newcommand{\eea}{\end{eqnarray}}
\newcommand{\be}{\begin{equation}}
\newcommand{\ee}{\end{equation}}
\newcommand{\ba}{\begin{align}}
\newcommand{\ea}{\end{align}}

\renewcommand{\epsilon}{\varepsilon}

   \makeatletter
  \let\over=\@@over \let\overwithdelims=\@@overwithdelims
  \let\atop=\@@atop \let\atopwithdelims=\@@atopwithdelims
  \let\above=\@@above \let\abovewithdelims=\@@abovewithdelims
\renewcommand\section{\@startsection {section}{1}{\z@}%
                                   {-3.5ex \@plus -1ex \@minus -.2ex}
                                   {2.3ex \@plus.2ex}%
                                   {\normalfont\large\bfseries}}

\renewcommand\subsection{\@startsection{subsection}{2}{\z@}%
                                     {-3.25ex\@plus -1ex \@minus -.2ex}%
                                     {1.5ex \@plus .2ex}%
                                     {\normalfont\bfseries}}

\begin{document}
\begin{titlepage}
\unitlength = 1mm
\ \\
\vskip 1cm
\begin{center}

{ \LARGE {\textsc{On BMS Invariance of Gravitational Scattering}}}

\vspace{0.8cm}
Andrew Strominger

\vspace{1cm}

{\it  Center for the Fundamental Laws of Nature, Harvard University,\\
Cambridge, MA 02138, USA}

\begin{abstract}
BMS$^+$ transformations act nontrivially on outgoing gravitational scattering data while preserving intrinsic  structure at
future null infinity (\ip). BMS$^-$ transformations  similarly act on ingoing data at past null infinity ($\ci^-$). In this paper we apply - within a suitable finite neighborhood of the Minkowski vacuum - results of Christodoulou and Klainerman to link \ip\ to $\ci^-$ and thereby identify ``diagonal" elements BMS$^0$ of BMS$^+\times $BMS$^-$.  We argue that BMS$^0$ is a nontrivial  infinite-dimensional symmetry of both classical gravitational scattering and the quantum gravity $\cs$-matrix. It implies the conservation of net accumulated energy flux at every angle on the conformal $S^2$ at $\ci$. The associated Ward identity is shown to relate $\cs$-matrix elements with and without soft gravitons. Finally,  BMS$^0$ is recast as a $U(1)$ Kac-Moody 
symmetry and an expression for the Kac-Moody current is given in terms of a certain soft graviton operator on the boundary of $\ci$.
 \end{abstract}

\vspace{1.0cm}

\end{center}

\end{titlepage}

\pagestyle{empty}
\pagestyle{plain}

\def\vx{{\vec x}}
\def\p{\partial}
\def\po{$\cal P_O$}

\pagenumbering{arabic}

\tableofcontents
\section{Introduction}

  BMS$^+$ transformations\footnote{The + superscript is added to denote the action is on \ip.} \cite{bms} are an infinite-dimensional set of ``large" diffeomorphisms which transform one asymptotically flat  solution of the general relativity constraints at future null infinity (\ip )  to a new, physically inequivalent solution. The infinite-dimensional supertranslation subgroup generates arbitrary angle (but not time) dependent translations of retarded time. Supertranslations for example relate solutions which have pulses of gravitational radiation emerging at different relative values of retarded times.  At the classical level BMS$^+$ transformations preserve the symplectic structure on \ip\ \cite{ash}. Quantum mechanically outgoing states at \ip\ (where gravity is weakly coupled)  are in representations of hermitian norm-preserving BMS$^+$ generators \cite{ash}.   There is an isomorphic structure at $\ci^-$ where ingoing states 
  are acted on by a second copy of the group denoted BMS$^-$.
  
In asymptotically Minkoswkian quantum gravity one seeks an $\cs$-matrix relating the in and out Hilbert spaces
\be |out\>=\cs |in\>.\ee
It is natural to ask whether or not BMS in any sense provides a symmetry of the $\cs$-matrix. An obvious guess would be 
a relation of the form
\be \label{ut}
B^{+}\cs-\cs B^-=0,
\ee
where $B^\pm$ are infinitesimal generators of BMS$^\pm$.
The classical limit of such a relation would give symmetries of classical gravitational scattering. 
Perhaps surprisingly, the existence of such a symmetry has not, to the best of our knowledge,  been pursued in the literature, for several reasons. The most important reason is that for a general asymptotically flat  space there is no canonical relation between BMS$^+$ and BMS$^-$  transformations  and hence no obvious choice of a specific $B^+$ to associate to a given $B^-$ in (\ref{ut}). Spatial infinity 
($i^0$) is a singular point in the conformal compactification of general asymptotically flat  spaces, preventing a fully general canonical identification between the far past of \ip\ and the far future of $\ci^-$. Clearly, without such an identification no relation of the form (\ref{ut}) can make sense. Further obstacles were complications due to black hole formation at the classical level and potential unitarity violation at the quantum level. 

In this paper we argue that in a finite neighborhood of the Minkowski vacuum classical gravitational scattering is in fact BMS-invariant, and that the perturbative quantum gravity $\cs$-matrix - assuming it exists \footnote{Hopefully troublesome IR divergences \cite{steve,stevebook} can be eliminated with the construction of \cite{Kulish:1970ut, Ware:2013zja}.}- obeys the corresponding relation (\ref{ut}). Most of these symmetries 
are spontaneously broken in the conventional Minkowski vacuum, $i.e.$  $B^\pm |0_M\>\neq 0$. Our argument relies heavily on the seminal work of Christodoulou and Klainerman \cite{ck} (henceforth CK) which appeared subsequent to the earlier investigations of \cite{bms, ash}.  We use the results of CK to show that, in a suitably defined finite neighborhood of the vacuum, $i^0$ is just smooth enough to allow a canonical identification between elements of $B^+$ and $B^-$, and hence ``diagonal' generators $B^0$ of  BMS$^+\times $BMS$^-$.
Morevoer we conjecture that, with this identification, (\ref{ut})  holds and $B^0$ generates exact symmetries of the nonperturbative $\cs$-matrix.

The diagonal generators are roughly comprised of Lorentz and supertranslation generators. From the Lorentz generators of $B^0$ we recover the usual Lorentz invariance of the $\cs$-matrix. The diagonal supertranslations  generate translations along the null generators of $\ci $\footnote{These generators are continued from 
$\ci^-$ to \ip\ thorugh $i^0$ using the standard conformal compactification: see section 2.4.} whose magnitude is constant along the generators  but an arbitrary function on the transverse conformal two-sphere.  As usual, symmetries imply conservation laws. Supertranslation invariance of the $\cs$ matrix in the form (\ref{ut}) implies that, for CK-type configurations which begin and end in the vacuum:  
\vskip .2cm
\noindent{\it The total incoming energy flux integrated along any null generator on $\ci^-$ equals the total outgoing energy flux integrated along the continuation of this null generator on \ip. }
\vskip .2cm
\noindent  In other words, local energy\footnote{  In general relativity the Bondi mass aspect $m_B$ defines this local energy at each angle on the conformal sphere at $\ci$.}
 is conserved at every angle.
 
At first this sounds nonsensical. A single free massless particle in Minkowski space which comes in from $\ci^-$ and goes out at \ip\ without changing propagation angle obviously conserves this local energy.  However a pair of interacting massless particles incoming from $\ci^-$  can for example scatter and go out to \ip\ at right angles.  So the particle  contributions to the local energies cannot in this case be conserved at every angle.  However, what happens is that soft gravitons are created which have zero $total$ energy, but have localized contributions to the energy which may be positive or negative, and are distributed in just such a manner so as to insure energy conservation at every angle.  In fact, the soft graviton contribution to the local Bondi energy is in some sense determined by requiring this to be true.  So on further reflection, the result may appear, rather than nonsensical,  to be a nearly trivial artifact of the Bondi definition of local energy  on $\ci$ in general relativity.    

Local energy conservation might have been anticipated as follows. Global time translations are supertranslations whose magnitude does not depend on the angle or position $(z,\bz)$ on the $S^2$.  These of course lead to global energy conservation. It is then plausible that the supertranslations which act at only one angle  lead to the conservation of energy  at that one angle. This indeed turns out to be the case. 

Even more interesting consequences of  supertranslation symmetry arise when applied to quantum scattering, where 
in and out states are not ordinarily in eigenstates of accumulated energy flux at each angle. 
The conservation law (\ref{ut}) can be rewritten in a Fock basis for the $\cs$-matrix. The explicit formula is a Ward identity relating any given scattering amplitude to the same amplitude with a soft graviton insertion. The latter arises because the supertranslation generators do not annihilate the usual Minkowski vacuum.  In a companion paper \cite{hms} we will show that this Ward identity
is precisely Weinberg's soft graviton theorem. 

We will also show that supertranslation Ward identities can be recast as Ward identity for a $U(1)$ Kac-Moody symmetry living on the conformal sphere at $\ci$. The Kac-Moody currents are constructed from soft gravitons living on the conformal $S^2$ at the boundary of $\ci$. This construction was motivated by the conjecture of \cite{bt, Banks:2003vp} that this $S^2$ admits not just an $SL(2,C)$ but a full Virasoro action, which is compatible with a reformulation of this type. 

Our discussion parallels recent analyses of the asymptotic/infrared structure of massless QED and Yang-Mills theory appearing in \cite{bal, juan,as}. The new methodology of the present paper may be useful in the gauge theory context. Related recent work on BMS  appears in \cite{Barnich:2013sxa, Barnich:2013axa}.

This paper is organized as follows. In section 2 we present our conventions and describe the configurations we wish 
to analyze:  weakly gravitating  CK geometries which begin and end in the vacuum. The BMS$^\pm$ actions are described and the diagonal generators identified. In section 3 we state and demonstrate BMS invariance of the $\cs$-matrix, show that the  supertranslation invariance implies energy conservation at every angle and derive a Ward identity relating $\cs$-matrix elements with and without soft gravitons. In section 4 we construct a two-dimensional $U(1)$ Kac-Moody current $P_z$ sourced by the net accumulated radiative energy flux at each angle through $\ci$. Finally we show that Kac-Moody Ward identities are equivalent to those following from supertranslations. 

\section{Vacuum-to-vacuum geometries}
 This paper is restricted to the analysis of  weakly gravitating scattering geometries which begin and end in the vacuum. In this section we give our conventions and characterize the spaces under consideration. 
\subsection{Asymptotically flat metrics} The structure of asymptotically Minkowskian geometries near \ip\ has been studied in great detail, see e.g. \cite{Wald:1984rg} and references therein.  A general Lorentzian metric can locally be written in retarded Bondi coordinates 
\begin{equation}
  \label{eq:2}
  ds^2=-U du^2-2e^{2\beta}dudr+
g_{AB}(dx^A-U^A du)(dx^B-U^B du)\,,
\end{equation}
with $\p_r {\rm det}{g_{AB} \over r^2}=0$. Here  $A,B=z,\bz$ and surfaces of constant retarded time $u$ are null.
Asymptotically flat metrics have an expansion around future null infinity ($r=\infty$) whose first few terms are
\bea \label{mt}~~~~~~~~~~~~~~ds^2& =& -du^2-2dudr+2r^2\gamma_{z\bz}dzd\bz \cr
~~~&~&+{2m_B \over r}du^2+rC_{zz} dz^2+r C_{\bz\bz}d\bz^2 -2U_z dudz - 2U_\bz dud\bz  \cr&&+ ....~~~,\eea
where \be \gamma_{z\bz} ={2\over (1+z\bz)^2},\ee 
is the round metric on the unit $S^2$,
\be \label{vv} U_z=-\half  D^z C_{zz}\ee and the indicated subleading corrections are suppress by inverse powers of $r$ relative to those written. The first line in (\ref{mt}) is the flat Minkowski metric. In the second line, $m_B(u,z,\bz)$ is the Bondi mass aspect,  which defines the local energy at retarded time $u$ at the angle on the asymptotic $S^2$ denoted $(z,\bz)$.  $D_z$ is the $\gamma$-covariant derivative\footnote{With our conventions  $\Gamma^z_{zz}=-2\bz/(1+z\bz),~R_{z\bz}=\gamma_{z\bz}$ and $[D_\bz,D_z]X^z=X_\bz$.} and here and hereafter $(z,\bz)$ indices are raised and lowered with $\gamma_{z\bz}$. The Bondi news $N_{zz}$, which characterizes the outgoing gravitational radiation, is
\be \label{nc} N_{zz}=\p_uC_{zz}.\ee
\ip\ is the null surface $(r=\infty, u, z, \bz)$. We use the symbol $\ci^+_+$ ($\ci^+_-$)  to denote the future (past) boundary  of 
\ip\ at $(r=\infty, u=\infty, z, \bz)$ (  $(r=\infty, u=-\infty, z, \bz)$). This is depicted in figure 1. 
\begin{figure}
\centering
\includegraphics[width=1.0\textwidth]{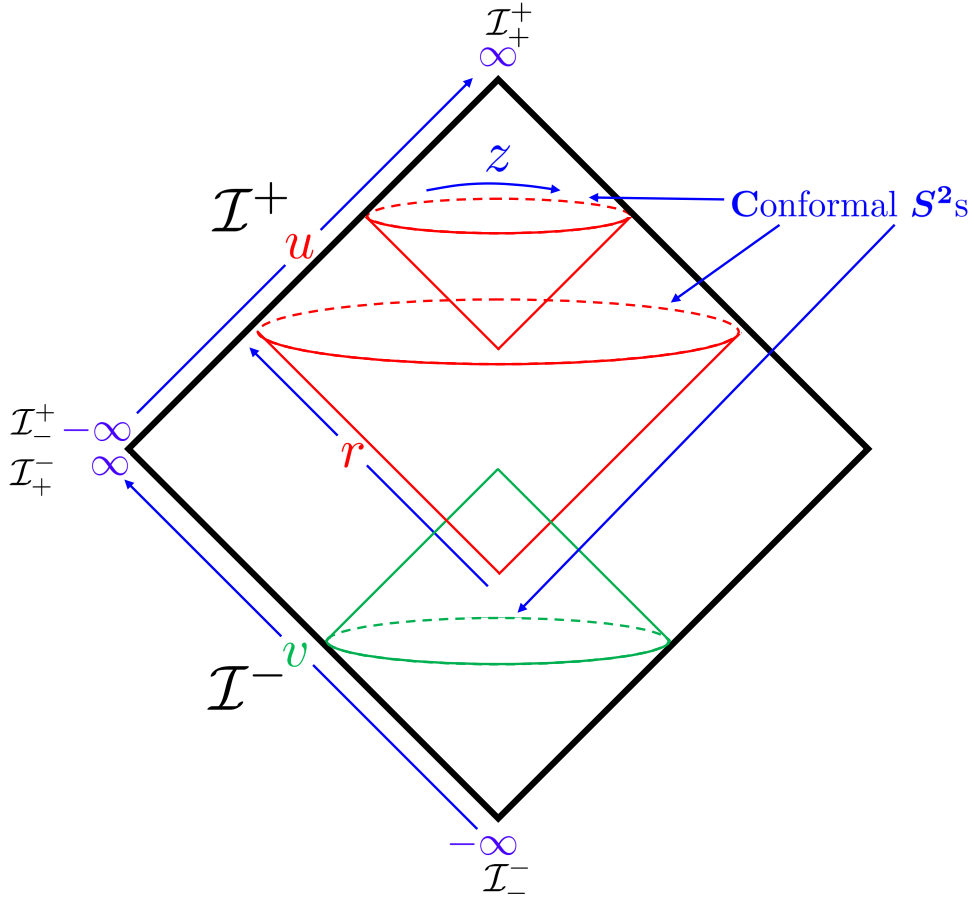}
 \caption{Penrose diagram for Minkowski space.  Near \ip\ surfaces of constant retarded time $u$ (red) are cone-like and intersect \ip\ in a conformal $S^2$ parametrized by $(z,\bz)$.   Cone-like surfaces of constant advanced  time $v$ (green) intersect $\ci^-$ in a conformal $S^2$ also parametrized by $(z,\bz)$. The future (past) $S^2$ boundary of \ip\ is labelled $\ci^+_+$ ($\ci^+_-$), while the future (past) boundary of  $\ci^-$ is labelled $\ci^-_+$ ($\ci^-_-$).}
\end{figure}
 
We are interested in geometries which are also asymptotically flat  at past null infinity ($\ci^-$ ).  This region may be described in the advanced Bondi coordinates
 \be \label{eq:3}
  ds^2=-V dv^2+2e^{2\beta^-}dvdr+
g^-_{AB}(dx^A-V^A dv)(dx^B-V^B dv)\,.
\end{equation}
where 
\be v(r,u,z,\bz)=u+2r+.... \ee
This has  the large $r$  expansion  
\bea \label{mpt}~~~~~~~~~~~~~~ds^2& =& -dv^2+2dvdr+2r^2\gamma_{z\bz}dzd\bz \cr
~~~&~&+{2m^-_B \over r}dv^2+rD_{zz} dz^2+r D_{\bz\bz}d\bz^2 -2V_z dvdz -2V_\bz dvd\bz  \cr&&+ ....~~~,\eea
where now
\be \label{bb} V_z=\half  D^z D_{zz}\ee
  $\ci^-$ is then the surface at  $r=\infty$  parametrized by $(v,z,\bz)$. $\ci^-_+$ ($\ci^-_-$) denotes the future (past) boundary  of 
$\ci^-$ at $(r=\infty, v=\infty, z, \bz)$ (  $(r=\infty, v=-\infty, z, \bz)$).  

We are interested in relating $\ci^-$ data - classical incoming gravity waves or quantum states - to \ip\ data. 
\ip\ and $\ci^-$ meet near spatial infinity $i^0$.  However the
relationship between $\ci^-_+$,  $\ci^+_-$ and $i^0$ is quite subtle \cite{ash}. $i^0$ is in general a singular point in the conformal compactification of an asymptotically flat geometry, and the conformally rescaled metric is not differentiable unless it is flat Minkowski space \cite{ash}. Therefore it need not in general be the case that for example $m_B|_{\ci^+_-}=m^-_B|_{\ci^-_+}$,  since the singularities are approached from a different direction.  Such relations between data on $\ci^+_-$and $\ci^-_+$ do exist for geometries in a suitably defined finite neighborhood of Minkowski space as discussed in section 2.4. 
\subsection{BMS  transformations}

Vector fields generating BMS transformations are of two types: there are six with an asymptotic $SL(2,C)$  Lie bracket algebra, and infinitely many commuting supertranslations. 
  Global $SL(2,C)$ conformal transformations are generated on \ip\ by the real part of the complex vector fields 
  \bea \label{cnt} \zeta^a_{\rm conf}\p_a=(1-{u \over 2r} )\zeta^z\p_z-(1+{u  \over  r} ){D_z\zeta^zr\over 2}\p_r -{u\over 2 r}\gamma^{z\bz} D^2_z\zeta^z\p_\bz+{u\over 2 }D_z\zeta^z\p_u ,\eea
  where\footnote{It has been proposed \cite{bt,Banks:2003vp} that the BMS algebra should be extended by allowing $\zeta^z$ 
  to be an arbitrary conformal Killing vector.} $\zeta^z=1,z, z^2,i, iz,iz^2$.  Here and below we suppress  (in some cases metric-dependent) terms which are further subleading in ${1 \over r}$ and irrelevant to our analysis: see \cite{bt} for a recent treatment specifying these terms.   
The Lie derivative acts as
  \bea \label{sl2c}\cL_\zeta C_{zz}&=&\zeta^z\p_zC_{zz} +2\p_z\zeta^z C_{zz} +{D_z\zeta^z \over 2}(u\p_u-1)C_{zz},\cr  \cL_{\bar \zeta} C_{zz}&=&\bar \zeta^\bz \p_\bz C_{zz} +{D_\bz \bar \zeta^\bz \over 2}(u\p_u-1)C_{zz},\cr \cL_\zeta m_B&=&\big(\zeta^z\p_z+{u\over 2 }D_z\zeta^z\p_u+{3D_z\zeta^z\over 2})m_B  \cr&&~~
  +{u\over 2}\p_u\big( U_z \zeta^z-U^zD^2_z\zeta^z \big).\label{eef}\eea
  $SL(2,C)$ transformations on $\ci^-$ are generated by 
  \bea \label{cntb} \zeta^{-a}_{\rm conf}\p_a=(1+{v \over 2r} )\zeta^z\p_z-(1-{v  \over  r} ){D_z\zeta^zr\over 2}\p_r +{v\over 2 r}\gamma^{z\bz} D^2_z\zeta^z\p_\bz+{v\over 2 }D_z\zeta^z\p_v,\eea
for which
  \bea \label{sl2cb}\cL_{\zeta^-} D_{zz}&=&\zeta^{-z}\p_zD_{zz} +2\p_z\zeta^{-z} D_{zz} +{D_z\zeta^{-z} \over 2}(v\p_v-1)D_{zz},\cr  \cL_{\bar \zeta^-} D_{zz}&=&\bar \zeta^{-\bz} \p_\bz D_{zz} +{D_\bz \bar \zeta^{-\bz} \over 2}(v\p_v-1)D_{zz},\cr \cL_{\zeta^-} m^-_B&=&\big(\zeta^{-z}\p_z+{v\over 2 }D_z\zeta^{-z}\p_v+{3D_z\zeta^{-z}\over 2})m^-_B  \cr&&~~
  -{v\over 2}\p_v\big( V_z \zeta^{-z}-V^zD^2_z\zeta^z \big).\label{eef}\eea

 Supertranslations on \ip\ are generated by the vector fields
\be \label{stn} f\p_u -{1\over r}(D^\bz f\p_\bz+D^z f \p_z )
+D^zD_z f\p_r,~~~~f=f(z,\bz). \ee
One finds \be \label{dp} \cL_fC_{zz}=f\p_uC_{zz}-2D_z^2 f,\ee
\be \label{dps} \cL_fU_{z}=f\p_uU_{z}-\half D^zf\p_uC_{zz}+D^z D_z^2 f.\ee
We will also be interested in $\ci^-$ supertranslations 
\be \label{ims} f^-\p_v +{1 \over r}(D^\bz f^-\p_\bz+D^z f^- \p_z )- D^zD_z f^-\p_r, \ee
under which
\be \label{dpw} \cL_{f^-}D_{zz}=f^-\p_vD_{zz}+2D_z^2 f^-,\ee
\be \label{dpd} \cL_{f^-} V_{z}=f^-\p_u V_{z}+\half D^zf\p_uD_{zz}+D^z D_z^2 f^-.\ee
Notice that, in our conventions for the metric coefficients,  the inhomogeneous terms in the transformation laws for $U_z$ and $V_z$ have the same sign, while those of $C_{zz}$ and $D_{zz}$ are opposite, consistent with the minus sign which appears in (\ref{vv}) but not (\ref{bb}). 

So far we have not specified any connection between BMS$^+$ and BMS$^-$ transformations. 
\subsection{Christodoulou-Klainerman spaces}
	 We are interested in asymptotically flat solutions of the Einstein equation which revert to the vacuum in the  far past and future. In particular we want to remain below the threshold for black hole formation.  We will adopt the rigorous definition of such spaces given by Christodoulou and Klainerman \cite{ck} (CK) who also proved their global existence and  analyzed their asymptotic behavior. 

CK  studied asymptotically flat initial data in the center-of-mass frame on a maximal spacelike 
slice for which the Bach tensor $\e^{ijk}D^ {(3)}_jG^{(3)}_{kl}$ of the induced three-metric decays like $r^{-7/2}$ (or faster) at spatial infinity 
and the extrinsic curvature like $r^{-5/2}$. This implies that in normal coordinates about infinity the  leading part of the three-metric has the (conformally flat) Schwarzschild form, with corrections which decay like $r^{-3/2}$. CK showed that all such initial data which moreover 
satisfy a global smallness condition give rise to a global, i.e. geodesically complete, solution.  We will refer to these solutions (and their boosts out of the center-of-mass frame) as CK spaces.

The smallness condition is satisfied in a finite neighborhood of Minkowski space, so this result established the stability of Minkowski space. Moreover many asymptotic properties of CK spaces at null infinity were derived in detail, see \cite{Christodoulou:1991cr} for a summary. In the following we quote only the properties of these spaces needed for our analysis.

   The gravitational radiation flux is proportional to the square of the Bondi news $N_{zz}$ which,  for any finite energy data,  vanishes on the boundaries of \ip \be\label{cnd} N_{zz}|_{\ci^+_\pm}
=0. \ee  For CK spaces the falloff for $u\to \pm \infty$ is 
\be \label{ff} N_{zz}(u) \sim |u|^{-3/2},\ee
or faster.
We are interested in the Weyl curvature component $\Psz_2$ which in coordinates (\ref{eq:2}) is given by \bea\Psz_2(u,z,\bz)&\equiv& -\lim_{r \to \infty}({r}C_{uzr\bz}\gamma^{z\bz})
 \cr &=&-m_B+ {1 \over 4} C^{zz}N_{zz}-\half\gamma^{z\bz}(\p_\bz U_z- \p_z U_\bz).\eea For  center-of-mass CK spaces at $u=\infty$ 
\be \label{io}\Psz_2 |_{\ci^+_+}=0,\ee
while at $u=-\infty$
\be \label{iob}\Psz_2 |_{\ci^+_-}=-M,\ee where $M/G$ ($G$ being Newton's constant) is the ADM mass. 
The real part of these relations together with (\ref{cnd}) imply
\be\label{df}   m_B|_{\ci^+_+}=0,\ee
\be\label{dcz}m_B|_{\ci^+_-}=M.\ee
The imaginary parts  yields
\be \label{ras} \big[\p_\bz U_z- \p_z U_\bz\big]_{\ci^+_\pm}=0.\ee
Similarly on $\ci^-$
\be\label{dss} m^-_B|_{\ci^-_-}=\big[\p_\bz V_z- \p_z V_\bz\big]_{\ci^-_\pm}=\p_vD_{zz}|_{\ci^-_\pm}=0,\ee
and
\be \label{xc} m^-_B|_{\ci^-_+}=M.\ee
The falloff of the Bondi news is 
\be M_{zz} \equiv \p_vD_{zz} \sim |v|^{-3/2} \ee
or faster.

We wish to stress that these properties are neither trivial or obvious.  For example while it is intuitively plausible that the total integrated mass should decay to zero at \ipp\ for a sufficiently weak gravitational disturbance, given the nonlocal, nonpositive, gauge-variant nature of local energy in general relativity, it is not obvious that unintegrated $m_B$ itself should go to zero at every angle. Moreover the news tensor could fall off more slowly near the boundaries of $\ci$ and still give a finite total energy flux. Both of these facts are central in the following. 

Note that  (\ref{dcz}) and (\ref{xc}) are not invariant under the  $SL(2,C)$ Lorentz transformation law for $m_B$ in equation (\ref{sl2c}), reflecting the fact that CK work in the center of mass frame.  We will consider more general $SL(2,C)$ frames below.

In the following we consider not only pure gravity but also the possible coupling to any kind of massless matter which dissipates at late (early) times on \ip\ ($\ci^-$).  The CK analysis has not been fully generalized to this case, although there is no obvious reason analogs of (\ref{io})-(\ref{xc}) might not still pertain to a suitably defined neighborhood of the gravity+matter vacuum. In the absence of such a derivation (\ref{io})-(\ref{xc})  will simply be imposed,  in the matter case,  as restrictions on the solutions under consideration.  

\subsection{Linking \ip\ to $\ci^-$  near $i^0$}

The \ip\ data  $m_B$ and $C_{zz}$ are related by the  constraint equation\footnote{The $G_{uz}$ constraint equation does not constrain this data: it determines further subleading components of $g_{uz}$ in terms of $C_{zz}$, $m_B$ and the matter stress tensor.}
\be \label{ct} \p_u m_B =-\half \p_u\big[D^zU_z+D^\bz U_\bz  \big]-T_{uu},\ee
where 
\be \label{tuu} T_{uu}={1 \over 4} N_{zz} N^{zz} +4\pi G\lim_{r\to \infty}\big[ r^2 T^M_{uu}\big]  \ee
is the total outgoing radiation energy flux - gravity plus matter - at \ip, rescaled by $4\pi G$ for notational convenience. 

CK spaces, as usually presented, are not quite solutions of 
the classical gravitational scattering problem. The latter consist of a set of final data on \ip\ which evolve from some initial data on $\ci^-$. In going from a CK space to this data there is a physical ambiguity under BMS$^+\times$BMS$^-$ transformations which we must understand how to fix. In particular we must 
determine, given the Bondi news and matter radiation flux, the leading terms $m_B$ and $C_{zz}$ of the metric in coordinates (\ref{mt}) everywhere  on 
\ip. To achieve this we integrate the constraint (\ref{ct}) and (\ref{nc}) along a null generator of \ip, but initial conditions at $\ci^+_-$ are required.  For center-of-mass CK spaces one has initially  constant $m_B|_{\ci^+_-}=M$ according to (\ref{dcz}). 
Under an $SL(2,C)^+$ transformation, which  takes us out of the center-of-mass frame, $m_B|_{\ci^+_-}$ transforms as 
\be  \cL_\zeta m_B|_{\ci^+_-}=\big(\zeta^z\p_z+{3D_z\zeta^z\over 2})m_B|_{\ci^+_-}, \ee
and is no longer constant.

The initial value of $C_{zz}$ may be any solution of (\ref{ras}), which is equivalent to 
\be\label{czz} \big[D^2_\bz C_{zz}- D_z^2 C_{\bz\bz}\big]_{\ci^+_-}=0.\ee
The general solution of this is
\be C_{zz}(-\infty,z,\bz)=D_z^2C,\ee
for some real function $C$, which implies $U_z=-\half \p_z(C+D_zD^z C)$. It follows from (\ref{dp}) that we can set $C=C_{zz}=U_z=0$ with a BMS supertranslation $f=\half C$ on \ip,  but we will not do so as that would explicitly break the BMS symmetry we wish to study.

Similarly  on $\ci^-$, $m^-_B$ is determined by the constraint
 \be \label{vct} \p_v m^-_B =\half \p_v\big[D^zV_z+D^\bz V_\bz  \big]+T_{vv},\ee
where $T_{vv}$ is the total incoming radiation flux at $\ci^-$. This may be solved by imposing initial data at $\ci^-_+$ and integrating down to $\ci^-_-$.  We take $m^-_B|_{\ci^-_+}=M$ in accord with (\ref{xc}) or, out of the center-of-mass frame an $SL(2,C)^-$ tranformation thereof. The mass aspect transforms as 
\be  \cL_\zeta m^-_B|_{\ci^-_+}=\big(\zeta^{-z}\p_z+{3D_z\zeta^{-z}\over 2})m^-_B|_{\ci^-_+}. \ee
In addition we take 
\be D_{zz}(-\infty,z,\bz)=D_z^2D,\ee
for some real function $D$.

However we are not done describing the asymptotic geometry because we must relate the inital data at $\ci^-_+$ to that of $\ci^+_-$. The description we have given so far allows independent BMS$^+$ transformations of \ip\ and 
BMS$^-$ transformations of $\ci^-$. This cannot be physically correct because for example the $SL(2,C)^\pm$ subgroups  change the total energy and an incoming configuration must scatter to an outgoing one with the same energy. 

In order to relate the initial data for the constraint equations we must first relate the points on $\ci^-_+$ to those on $\ci^+_-$.
A natural relation derives from the conformal compactification of asymptotically Minkowskian spacetimes in which spatial infinity $i^0$ is a point and \ip\ ($\ci^-$) its future (past) lightcone. Null generators of $\ci$ then run from $\ci^-$ to \ip\ though $i^0$.  These generators give us the required identification: we label all points on the same generator by the same coordinate $(z,\bz)$. Note that this involves a kind antipodal map between \ip\ and $\ci^-$: a light ray crossing  the interior flat Minkowski space begins and ends at the same value of $(z,\bz)$. 

We first consider the subgroup $SL(2,C)^+\times SL(2,C)^-$, which cleary must be cut down to a smaller energy-conserving group. The desired $SL(2,C)^0$ subgroup is obviously the one which corresponds to ordinary Lorentz transformations for the case of Minkowski space. It is easy to see that these are generated by vector fields of the form (\ref{cnt}), (\ref{cntb}) restricted by the condition
\be \label{sds} \zeta^z=\zeta^{-z}.\ee
Note this this condition is not even defined before we have an identification of points on $\ci^-_+$ and $\ci^+_-$.  These transformations preserve the condition 
\be \label{szz} m_B|_{\ci^+_-}(z,\bz)=m_B|_{\ci^-_+}(z,\bz) \ee
although neither mass aspect remains constant under boosts. 

It remains to relate the initial data for $C_{zz}$ at $\ci^+_-$ to that for $D_{zz}$ at $\ci^-_+$ in a manner compatible with (\ref{sds}). This is accomplished by imposing\be \label{CTM}D_{zz}(\infty,z,\bz)=-C_{zz}(-\infty,z,\bz).\ee
(\ref{vv}) and (\ref{bb}) then immediately imply 
\be\label{rf} V_z|_{\ci^-_+}=U_z|_{\ci^+_-}.\ee
This condition then breaks the separate BMS symmetries of $\ci^+$
and $\ci^-$ down to a single one which acts on both 
\be \label{xz}
f^-(z,\bz)=f(z,\bz). \ee   The case $f=f^-=constant$ is global time translations. The three constant spatial translations are given by
\be \label{st} f_{st}={1-z\bz \over 1+z\bz} \ee
together with  spatial rotations thereof.

Generators obeying (\ref{sds}) and (\ref{xz})  are the diagonal generators of BMS$^+ \times$BMS$^-$ referred to in the introduction. These generators are the ones that are constant on the null generators of $\ci$ even as they pass through $i^0$. We emphasize that the linking of data on \ip\ and $\ci^-$ depends crucially on properties of CK spaces and may not be possible for more general types of asymptotically flat spaces.
For example if the news tensor fell off like $|u|^{-1}$, 
$C_{zz}$ would diverge logarithmically and the key boundary condition (\ref{CTM}) could not be imposed.

\subsection{Symmetries of classical gravitational scattering}

Let us now summarize the implications of diagonal BMS$^0$  invariance for classical gravitational scattering. 
Begin with a CK space in the center-of-mass frame, together with the conformally invariant radiative data $N_{zz}$ and $M_{zz}$ on $\ci$. To view this as scattering solution, we must construct both the initial data on $\ci^-$ and the final data on \ip. In order to do this, we take 
$ m_B|_{\ci^+_-}(z,\bz)=m_B|_{\ci^-_+}(z,\bz) =M$ and, for example, $ C_{zz}|_{\ci^+_-}(z,\bz)=D_{zz}|_{\ci^-_+}(z,\bz) =0$ and then integrate the constraint equations on both \ip\ and $\ci^-$ away from $i^0$.  This resulting  final data $(C_{zz}, m_B)$ is the solution of the scattering problem for the initial data $(D_{zz},m^-_B)$.

The diagonal BMS symmetry can now be used to generate new solutions of the scattering problem from this one. The simplest case are the diagonal $SL(2,C)$ generators. They simply tell us that, given a pair of initial and final data that solve the scattering problem, a new pair can be generated by boosting or rotating both the initial and final data sets. Similarly from the four special supertranslations, we learn that translated initial data scatters into translated final data. 
These statements are familiar and perhaps obvious. 

A less obvious statement derives from consideration of generic supertranslations. 
Suppose we send in gravity wave pulses from the north  and south pole both near advanced time $v=0$ and find that they scatter to outgoing wave pulses also at the north and south pole 
emerging near  $u=0$. Supertranslation symmetry can then be used to show that the supertranslated ingoing data with wave pulses from the north  pole at $v=0$ and south pole at say $v=7$ scatters to outgoing data  with wave pulses emerging near  the north  pole at $u=7$ and south pole $u=0$.

Hence for every CK space we can generate an infinite parameter BMS$^0$ family of solutions to the scattering problem. Some of these differ by the obvious Poincare tranformations, while others differ by the less obvious  supertranslations. 
\section{BMS invariance of the $\cs$-matrix}
   In this section we argue that the $\cs$-matrix is BMS-invariant, translating the classical symmetry described above into the semiclassical language of states and operators. We then derive a  Ward identity in a Fock basis. We concentrate largely on infinitesimal supertranslations, as Lorentz invariance is well-understood, and for simplicity work in the center-of-mass frame.  
   
   \subsection{Supertranslation generators commute with $\cs$}
In this subsection we describe the generators of  supertranslations on $\ci^\pm$ and show that they commute with $\cs$. 

Classical general relativity admits a simple  Hamiltonian formulation at \ip\ where gravity is weakly coupled. 
The phase space $\Gamma^+_R$ of radiative modes is parametrized by the Bondi news $N_{zz}$. The symplectic product on the tangent space of $\Gamma^+_R$ is the projection of \cite{ash}
\be \label{zn} {1 \over 16\pi G}\int_{\ci^+}du d^2z\gamma^{z\bz}\delta C_{z z}\overleftrightarrow {\p_u} \delta C_{\bz\bz}. \ee
This implies the Dirac bracket
\be\label{db}  \{N_{\bz\bz}(u,z,\bz),N_{ww}(u',w,\bw)\}=-16\pi G \p_u\delta(u-u')\delta^2(z-w)\gamma_{z\bz}.\ee
It is important in the following that $\Gamma^+_R$ does not capture all the structure of interest at \ip. In particular the zero mode $\int du \delta N_{zz}$ along a null generator has vanishing symplectic product with everything in, and hence  is not tangent to,  $\Gamma^+_R$.
These zero modes lead to the ``infrared sectors" discussed in \cite{ash}, and are related to soft gravitons. We leave a more complete discussion of this larger and very interesting  space to future investigations. 

The generator of BMS supertranslations on \ip\ is 
\be  T(f)={1 \over 4\pi G}\int_{\ci^+_-}d^2z \gamma_{z\bz}f(z,\bz)m_B.\ee Note that $T(1)=M$ is the ADM Hamiltonian (times $G$). 
Using the constraints this can be written as an \ip\ integral\footnote{We are dropping here a term proportional to the late time mass aspect $m_B|_{\ci^+_+}$ which vanishes for the classical solutions and semiclassical states considered herein, but could be important in other applications.}
\be T^+(f)= {1 \over 4\pi G}\int dud^2zf\big[ \gamma_{z\bz}T_{uu} +\half \p_u (\p_zU_\bz+\p_\bz U_z)\big].\ee
The first term generates supertranslations on the radiative phase space $\Gamma^+_R$, i.e,\cite{ash,bt}
\bea \label{cb}\{ {1 \over 4\pi G}\int_{\ci^+}du d^2z\gamma_{z\bz}fT_{uu},N_{zz}\}&= &f \p_uN_{zz},\cr 
 \{ {1 \over 4\pi G}\int_{\ci^+}du d^2z\gamma_{z\bz}fT_{uu},\Phi \}&=& f \p_u\Phi.\eea
where $\Phi$ denotes any massless matter field on \ip. 
The linear term in $T^+(f)$ can be written as a boundary term. Hence it commutes with $N_{zz}$ and acts trivially on $\Gamma^+_R$.\footnote{As will be shown in detail in \cite{hms}, the Dirac brackets of these boundary terms must account for (\ref{czz}) as a constraint on physical excitations. Once this is incorporated, $T^+$ as written also generates the inhomogeneous, $u$-independent,  term in the action of the supertranslations on $C_{zz}$ itself - which is not a coordinate on $\Gamma^+_R$. } 
To interpret this term let us write it as 
\be {1 \over 8\pi G} \int_{\ci^+} dud^2zf\p_u\big[ \p_zU_\bz+\p_\bz U_z]=-\lim_{\omega \to 0}{1\over 32\pi G}  \int_{\ci^+} dud^2z(e^{i\omega u}+e^{-i\omega u})\big[N_\bz^{~z} D^2_z f +N_z^{~\bz} D_\bz^2 f ].\ee 
In this form it is  recognizable as  the soft ($\omega \to 0$) limit  of a metric perturbation with polarization tensor proportional to $D_z^2f$. In the quantum theory this will become a soft graviton field operator.

We describe  this by saying that the action of $T^+(f)$ generates supertranslations on the radiative phase space $\Gamma^+_R$, and also adds a soft graviton. This is reminiscent of the 
action of the Virasoro generator $L_{-2}$ on a generic state of AdS$_3$ quantum gravity: it infinitesimally translates the bulk fields while adding a boundary graviton.  The soft graviton can be viewed as the Goldstone boson predicted by Goldstone's theorem: according to the classical transformation law (\ref{dp}), non-constant supertranslations do not leave the vacuum invariant.\footnote{I  am grateful to A. Zhiboedov for this point.}

One may also define:
\be  T^-(f)={1\over 4\pi G}\int_{\ci^-_+}d^2z \gamma_{z\bz}fm^-_B={1 \over 4\pi G}\int dvd^2zf\big[ \gamma_{z\bz}T_{vv} +\half \p_v (\p_zV_\bz+\p_\bz V_z)\big],\ee
which similarly generates supertranslations on $\ci^-$. These generators obey
\be\label{ss} \{T^+(f),T^+(f')\}=0.\ee

On CK spaces  in coordinates obeying (\ref{CTM}) and (\ref{szz})
one has 
\be \label{rr}T(f)=T^+(f)=T^-(f). \ee 
This has implications for the semiclassical $\cs$-matrix, whose elements we denote 
\be \<out|\cs|in \>.\ee 
Since $T(1)$ is the Hamiltonian, and $\cs$ is constructed from exponentials of the Hamiltonian, it follows from (\ref{ss}) that infinitesmal BMS transformations commute with the $\cs$-matrix $[T,\cs]=0$. 
In particular
\be  \<out|\big( T^+(f)\cs-\cs T^-(f)\big)|in \>=0.\ee
This expresses infinitesimal BMS$^0$ invariance of the $\cs$-matrix.

\subsection {Energy conservation at every angle}
The identity (\ref{rr}) for the special choice $f=\delta^2(z-w)$ reads
\be \label{ec} \int_{\ci^+} du\big[ \gamma_{z\bz}T_{uu} +\half\p_u (\p_zU_\bz+\p_\bz U_z)\big]=\int_{\ci^-} dv\big[  \gamma_{z\bz}T_{vv} +\half \p_v (\p_zV_\bz+\p_\bz V_z)\big].\ee
This means that the total accumulated energy incoming from every angle $(z,\bz)$ on $\ci^-$ equals the total accumulated energy emerging  at the angle $(z,\bz)$ on \ip.  This conservation law relies crucially on the linear soft-graviton contribution to the unintegrated Bondi energy $m_B(z,\bz)$. The soft graviton contribution  is a total derivative on the sphere  and so does not contribute to the total integrated Bondi energy. 


Using the continuity condition (\ref{rf}) together with (\ref{ras}) and (\ref{dss}), (\ref{ec}) can be rewritten
\be \label{xcr} \gamma_{z\bz}\big(\int_{\ci^+} duT_{uu}-\int_{\ci^-} dvT_{vv}\big)=\p_\bz V_z|^{\ci^-_+}_{\ci^-_-}-\p_\bz U_z|^{\ci^+_+}_{\ci^+_-}.\ee
Here we have eliminated $V_\bz$ and $U_\bz$ terms using (\ref{ras}), (\ref{dss}). Of course we could equally have eliminated $V_z$ and $U_z$: this point will be returned to in section 4.2.

\subsection{Fock representation of BMS symmetry}
So far our statements have  largely been in terms of classical fields or their semiclassical counterparts.  Now we would like to consider their realization in terms of a Fock basis of quantum particles. 

Our derivation of classical BMS invariance held for weakly gravitating CK spaces. Quantum scattering theory, typically described perturbatively by Feynman diagrams, is based on a different type
of field configuration - often plane waves. Hence it is not obvious that our derivation should apply to the $\cs$-matrix in a Fock basis. Nevertheless we shall simply assume that it does, and the results so derived appear reasonable. Ultimately, the justification lies in their equivalence to Weinberg's soft graviton theorem \cite{hms}.

Let us denote  an in-state comprised of particles with energies $E^{in}_k$ incoming at points $ z^{in}_k$ on the conformal $S^2$ by  
\be |z^{in}_1,z^{in}_2,...\>. \ee
The semiclassical supertranslation operator then obeys the quantum relation \be T^-(f)|z^{in}_1,z^{in}_2,...\>  = F^-|z^{in}_1,z^{in}_2,...\>+\sum_k E^{in}_k f(z_k)  |z^{in}_1,z^{in}_2,...\> ,  \ee
where $F^-$ denotes the incoming soft graviton operator with polarization tensor proportional to $D_z^2f$: 
\be F^- ={1 \over 4\pi G} \int_{\ci^-}dv d^2z f \p_v \p_\bz V_z= { 1\over 8 \pi G} \int_{\ci^-} dvd^2z D_\bz^2f  M^\bz_{~z} ,\ee where $M_{zz}=\p_vD_{zz}$ is the news tensor on $\ci^-$.
We also have 
\be \<z^{out}_1,z^{out}_2,...|   T^+(f)=  \< z^{out}_1,z^{out}_2,...|F^+ +\sum_k E^{out}_k f(z^{out}_k)  \<z^{out}_1,z^{out}_2,...| ,\ee
with the outgoing soft graviton operator defined as 
\be  F^+= { 1\over 4 \pi G} \int_{\ci^+} dud^2z f  \p_u\p_\bz U_z=- { 1\over 8 \pi G} \int_{\ci^+} dud^2z D_\bz^2f  N^\bz_{~z}.\ee
Let us write \be \label{fpm} F\equiv F^+-F^- = -{ 1\over 8 \pi G} \int d^2z \gamma^{z\bz}D_\bz^2f  \big[ \int_{\ci^-} dv M_{zz}+\int_{\ci^+}du N_{zz}\big].\ee 
The square brackets is the  zero-mode of the news tensor along a null generator of $\ci$            which runs from $\ci^-$ to \ip\ through $i^0$. Hence it is a soft graviton. 
Adopting  the notation $:...:$  to denote the time ordered product, so that \be :FS:=F^+\cs-\cs F^-,\ee
BMS invariance of the $\cs$-matrix  implies 
\be\label{gg}  \<z^{out}_1,z^{out}_2,...|:F\cs: |z^{in}_1,z^{in}_2,...\>=  \sum_k \big(E^{in}_k f(z^{in}_k)-E^{out}_k f(z^{out}_k)\big)  \<z^{out}_1,z^{out}_2,...|\cs|z^{in}_1,z^{in}_2,...\>.\ee
This is a supertranslation Ward identity relating  $\cs$-matrix elements with and without an insertion of a soft graviton with polarization $D_z^2f$. In the companion paper \cite{hms} this relation is identified with Weinberg's soft graviton theorem. 

It follows immediately from (\ref{dp}) and (\ref{dpw}) that the soft graviton insertions considered here are pure gauge. 
However the required gauge transformations are supertranslations which are supposed to be  nontrivial elements of the asymptotic symmetry group. Indeed,  we here verify that they are nontrivial from the fact that they have non-vanishing Fock basis $\cs$-matrix elements. 
\section{Supertranslations as a Kac-Moody symmetry}

In this section we show that the supertranslation generators of the previous section can be recast as those of  a non-compact Kac-Moody algebra  acting on the asymptotic $S^2$ at $\ci$. This is in harmony with the proposal of \cite{bt, Banks:2003vp}, that this $S^2$ admits a Virasoro action.  
\subsection{The classical supertranslation  current}
In this subsection we construct  a classical current on the conformal $S^2$ at null infinity that is locally sourced by the difference at a given  point $(z,\bz)$ on $S^2$ of the outgoing and incoming energy fluxes. Define
 \be P_z= {1 \over 2 G}\big(V_z|^{\ci^-_+}_{\ci^-_-}- U_z|^{\ci^+_+}_{\ci^+_-} \big)   ,\ee
which is just $F$ for the case $f={1 \over z-w}$.
 Then (\ref{xcr}) becomes 
 \be \p_\bz P_z={\gamma_{z\bz}\over 2 G}\big( \int_{\ci^+} duT_{uu}-\int_{\ci^-} dvT_{vv}\big).\ee
 The solution of this is 
\be  P_z={1 \over 4\pi G   }\int d^2 w { \gamma_{w \bw} \over z-w} \big(\int du  T_{uu }-\int dv T_{vv}\big). \ee 
Consider the special case in which the Bondi news and matter radiation carry energy $E^{out}_k$ ($E^{in}_k$) across \ip\  ($\ci^-$) at discrete points $(u_k,z^{out}_k)$ ($(v_k,z^{in}_k)$with \be G\sum _kE^{out}_k=  G\sum _kE^{in}_k=M.\ee
We may then approximate near \ip\
\be  T_{uu}= 4\pi G\sum_kE^{out}_k\delta(u-u_k){\delta^2(z-z^{out}_k)\over \gamma_{z\bz}} ,\ee
while near $\ci^-$ \be  T_{vv}= 4\pi G\sum_kE^{in}_k\delta(v-v_k){\delta^2(z-z^{in}_k)\over \gamma_{z\bz}}~.  \ee 
We then find 
\be \label{cnkb}  P_z= \sum_k  \big( {E^{out}_k\over z-z^{out}_k}-{E^{in}_k\over z-z^{in}_k}\big)~.\ee
This is the classical version of the Ward identity for a $U(1)$ Kac-Moody current. 

\subsection{Kac-Moody Ward identities}
This subsection derives the semiclassical  version of (\ref{cnkb}) for the  classical current $P_z$
and shows that contour insertions of the quantum current generate the Kac-Moody symmetry.

Consider the supertranslation which acts on only one null generator of $\ci$: \be \label{dg} f (z,\bz)=\delta^2(z-w).\ee
Then the Ward identity (\ref{gg}) becomes 
\bea\label{ggz}  &&\p_\bw\<z^{out}_1,z^{out}_2,...|:P_w\cs: |z^{in}_1,z^{in}_2,...\>\cr &&~~~~~~~~= 2\pi \sum_k \big(E^{out}_k \delta^2(w-z^{out}_k)-E^{in}_k \delta^2(w-z^{in}_k)\big)  \<z^{out}_1,z^{out}_2,...|\cs|z^{in}_1,z^{in}_2,...\>, ~~~~~\eea
which in turn implies 
\be\label{ggs}  \<z^{out}_1,z^{out}_2,...|:P_w\cs:| z^{in}_1,z^{in}_2,...\>=  \sum_k \big({E^{out}_k \over w-z^{out}_k}-{E^{in}_k \over w-z^{in}_k}\big)  \<z^{out}_1,z^{out}_2,...|\cs|z^{in}_1,z^{in}_2,...\>, \ee
where we have used the fact that spatial differentiation $\p_\bw$ commutes with time ordering.
This is the semiclassical version of the classical equation (\ref{cnkb}).
It states  that the supertranslation Ward identity governing insertions of the current $P_z$ is exactly that of a $U(1)$ Kac Moody current whose charge is 
 the net accumulated energy flux across a point in the conformal $S^2$ at $\ci$.

Define the operator
\be P_\e ={1 \over 2 \pi i}\oint_C\e P_z,\ee
where the contour $C$ surrounds all the points $(z^{in}_k,z^{out}_k)$ and $\e$ is holomorhpic in this region.
Then 
\be\label{ggs}  \<z^{out}_1,z^{out}_2,...|:P_\e \cs:| z^{in}_1,z^{in}_2,...\>=  \sum_{k} \big({E^{out}_k \e(z^{out}_k)}-{E^{in}_k \e (z^{in}_k)}\big)  \<z^{out}_1,z^{out}_2,...|\cs|z^{in}_1,z^{in}_2,...\>. \ee
Hence we see that insertions of contour integrals of the soft graviton current $P_z$ generate a 
noncompact $U(1)$ supertranslation Kac-Moody algebra. 

Is the Kac-Moody symmetry equivalent to the supertranslation symmetry? Given that an
insertion of the the derivative of the Kac-Moody current  at $w$ has the same effect as a supertranslation with $f(z)=\delta^2(z-w)$ as in (\ref{dg}), it is clear that the Kac-Moody Ward identities associated to the insertions of the current $P_z$  capture all the information associated to supertranslation invariance -- at least concerning $\cs$ matrix identities. However the information is packaged differently: $P_z$ is $not$ the generator  of supertranslation on the incoming or outgoing Hilbert spaces. 

What about insertions of the antiholomorphic current $P_\bz$? These correspond to negative rather than positive helicity soft gravitons. Derivatives of such insertions 
are also equivalent to the choice $f(z)=\delta^2(z-w)$. Therefore they  do not seem to lead to any new information, at least in the form of $\cs$-matrix identities in a Fock basis. An equivalent statement  is that the difference of a negative and positive helicity soft graviton insertion with asymptotic polarization tensors $D_z^2f$ and $D_\bz^2f$, for any real $f$, vanishes in $\cs$-matrix elements and should be regarded as trivial. Hence we are not missing anything by considering only positive helicity gravitons. This leads us to suspect the  asymptotic symmetry transformations, whose elements by definition act nontrivially, is generated by only one current, which we will take to be  the holomorphic one $P_z$. 
While we have not ruled out some other use of the antiholomorphic current $P_\bz$,  inclusion of the a $P_\bz$ current algebra would appear to be redundant.  This suggests that the asymptotic symmetry should be thought of as a single $U(1)$ Kac Moody symmetry on the conformal $S^2$  at $\ci$.

\section*{Acknowledgements}
I am grateful to A. Ashtekar, T. Banks,  J. Bourjailly, D. Christodoulou, G. Compere, T. He, G. Horowitz, V.  Lysov,  P. Mitra, G. S. Ng and A. Zhiboedov for useful conversations and especially to J. Maldacena for critical insights at several stages of this project.  This work was supported in part by NSF grant 1205550  and the Fundamental Laws Initiative at Harvard.

\end{document}